# Coarse-grained model of the J-integral of carbon nanotube reinforced polymer composites

Behrouz Arash[3,*], Harold S. Park[4], Timon Rabczuk[1, 2, 3, 5,†]

[1]*Division of Computational Mechanics, Ton Duc Thang University, Ho Chi Minh City, Vietnam.*
[2]*Faculty of Civil Engineering, Ton Duc Thang University, Ho Chi Minh City, Vietnam*
[3]*Institute of Structural Mechanics, Bauhaus Universität-Weimar, Marienstr 15, D-99423 Weimar, Germany*
[4]*Department of Mechanical Engineering, Boston University, Boston, Massachusetts 02215, USA*
[5]*School of Civil, Environmental and Architectural Engineering, Korea University, Seoul, Republic of Korea*

**Abstract**

The J-integral is recognized as a fundamental parameter in fracture mechanics that characterizes the inherent resistance of materials to crack growth. However, the conventional methods to calculate the J-integral, which require knowledge of the exact position of a crack tip and the continuum fields around it, are unable to precisely measure the J-integral of polymer composites at the nanoscale. This work aims to propose an effective calculation method based on coarse-grained (CG) simulations for predicting the J-integral of carbon nanotube (CNT)/polymer composites. In the proposed approach, the J-integral is determined from the load displacement curve of a single specimen. The distinguishing feature of the method is the calculation of J-integral without need of information about the crack tip, which makes it applicable to complex polymer systems. The effects of the CNT weight fraction and covalent cross-links between the polymer matrix and nanotubes, and polymer chains on the fracture behavior of the composites are studied in detail. The dependence of the J-integral on the crack length and the size of representative volume element (RVE) is also explored.

---

[*] Author to whom correspondence should be addressed. E-mail address: behrouz.arash@uni-weimar.de, Tel: +49 3643 584511.

[†] Author to whom correspondence should be addressed. E-mail address: timon.rabczuk@tdt.edu.vn, Tel: +49 3643 584511.

## 1. Introduction

As a developing class of composite materials, short-fiber-reinforced polymers (SFRPs) have attracted much attention because of their remarkable mechanical, thermal and electrical properties [1-3]. The compromise between relatively low cost and high performance have made the composites excellent alternatives in various science and engineering applications such as electronic, automotive, oilfield and chemical industries [4]. In order to develop SFRPs, studies on failure mechanisms in the materials are of great significance [5, 6]. In classical fracture mechanics, the J-integral [7] is used to quantify the energy release rate in an elastic-plastic body that contains a crack. The J-integral has been widely used as a fracture characterizing parameter in macroscale mechanics to measure materials inherent resistance to crack growth [8].

In nanoscience and nanoengineering, where sizes are scaled down to nanometers, materials can no longer be considered as continua, but discrete molecular structures. It is, therefore, interesting to examine that if the fracture parameter developed based on the continuum theory is applicable to the atomistic structures. In order to address this issue, a number of studies have been conducted to extend the concept of J-integral to the nanoscale for the analysis of nano-sized cracks and their influence on the performance of materials [9-13]. Jin and Yuan [14] calculated the energy release rate of edge-cracked and center-cracked graphene sheets (GSs) in mode I and II fracture using molecular dynamics (MD) simulations and continuum mechanics. They proposed two methods for calculating the energy release rate. The first one is based on the variation of total potential energy of the same GSs with different crack lengths during the crack growth. This technique is computationally expensive, which makes it infeasible for large atomic systems. In the second approach, the energy release rate is calculated using the local forces at the crack tip. Although the latter method is less time consuming, the exact position of the crack tip is required for calculating the energy release rate. This limitation restricts the scheme to study the fracture behavior of crystal materials in which the crack propagation is uniform and the crack tip can be clearly identified. Another restriction of the methods is that they are restricted to linear elastic fracture mechanics (LEFM). Khare et al. [15] developed a coupled quantum mechanical/molecular mechanical model to study the effects of large defects and cracks on the elastic-plastic fracture behavior of carbon nanotubes (CNTs) and GSs. They used a domain integral approach for calculating the J-integral. The application of the method involves calculating the strain energy density, the traction vector and displacement vector components on a contour around the tip of a crack. The integral method is, however, complicated to be used for amorphous polymer materials, since crack paths in polymers and their composites can be irregular at the nanoscale. This causes difficulties in exactly identifying the crack tip position, and so the calculation of the continuum fields around the crack tip is computationally challenging in polymer systems. Xu et al. [16] presented a J-integral calculation method using the energy release rate during crack growth by which they explored the ductile fracture behavior of nickel crystals. This approach is limited to cracks that stably propagate because of the need for a finite difference approximation of the variation of potential energy with respect to the crack length. Recently, a methodology for calculating the J-integral using the continuum fields of stress, energy density and displacement gradient obtained from MD simulations, and then integrating the resulting Eshelby stress over a contour surrounding a crack tip was developed by Jones and Zimmerman [17]. The method was applied to quasi-static calculations to determine the J-integral of face-centered cubic and body-centered cubic crystals at zero and finite temperatures [18, 19]. Results attained by this method were shown to be in agreement with those predicted by LEFM. Despite the successful use of the proposed method in predicting the J-integral of crystal materials, the need for continuum fields around a crack tip and the exact position of the crack tip hinders its application to amorphous polymer materials. As formerly mentioned, crack propagation in polymer materials is unstable, and crack branches and kinks may occur in the crack path. Hence, obtaining the continuum fields from MD simulations and mapping them to a continuum framework are not feasible for polymer materials in practice.

In view of the abovementioned problems, there is still no accurate elastic-plastic J-integral calculation method for cracks in amorphous polymers and their composites. Furthermore, the effects of nanotube reinforcements on the J-integral of short CNT/polymer composites have not been quantified. Hence, a simulation method for predicting the J-integral of reinforced polymer composites and a quantitative study on the fracture behavior of the composites are essential to achieve a successful design, synthesis, and characterization of the nanocomposites.

In this article, a simulation method for calculating the J-integral of reinforced polymer composites is proposed. The present method is based on the determination of the J-integral from the load displacement curve of a single specimen without need of information about the position of a crack tip and local continuum fields around the tip. The characteristic feature makes the approach feasible to complex polymer composites. A coarse-grained (CG) model [20, 21] for investigating the elastic-plastic fracture behavior of CNT/poly (methyl methacrylate) (PMMA) polymer composites is developed. The principle of CG models is to map a set of atoms to a CG bead, which enables to extend the accessible time and length-scales while maintaining the molecular details of an atomistic system. The effects of the nanotube weight fraction, covalent cross-links in the composites and the size of representative volume element (RVE) on the fracture characteristics of CNT/PMMA composites are studied in detail.

## 2. Coarse-grained model

Up to now, many CG models have been developed for CNTs and polymer materials in the literature [22-24]. In this study, a CG model that was successfully examined in predicting the elastic-plastic properties of CNT/PMMA composites [25, 26] is used. In the CG model, each monomer of methyl methacrylate ($C_5O_2H_8$) is mapped into a CG bead hereafter named P bead with an atomic mass of 100.12 amu as illustrated in Fig.1a. The center of the bead is chosen to be the center of mass of the monomer. Each five atomic rings of (5, 5) CNTs is treated as a CG bead with an atomic mass of 600.55 amu defined by C bead as shown in Fig.1b. The center of C beads is assumed to be the center of the five atomic rings. The degree of freedom reduction is 15 and 50 folds for P and C beads, respectively, as compared to full atomistic systems.

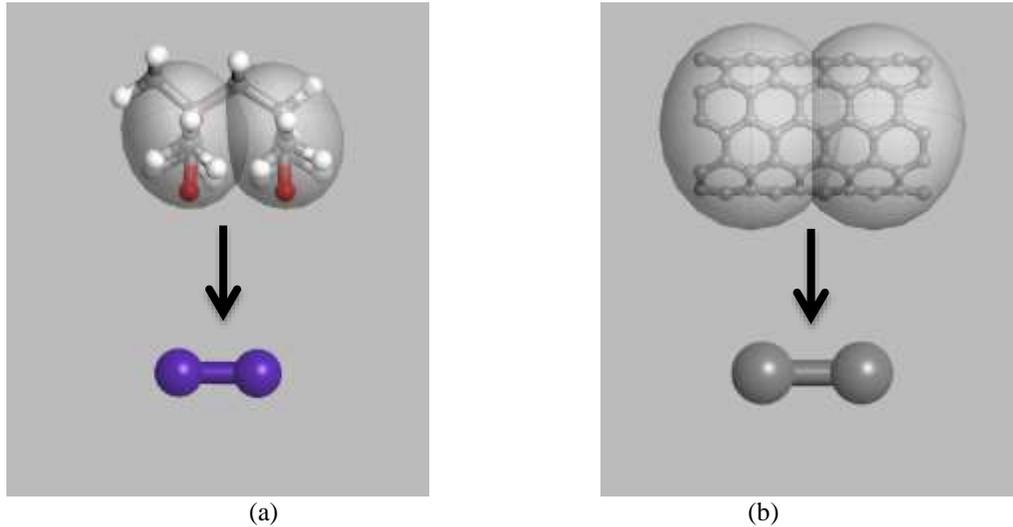

(a) (b)

Fig. 1. (a) Two monomers of a PMMA polymer chain and its CG model made of two P beads, and (b) a (5, 5) CNT with 10 rings of carbon atoms and its CG model made of two C beads.

The total potential energy, $E_{total}$, of a system can be written as the sum of energy terms associated with the variation of the bond length, $E_b$, the bond angle, $E_a$, the dihedral angle, $E_d$, the vdW interactions, $E_{vdW}$, as $E_{total} = \sum_i E_{b_i} + \sum_j E_{a_j} + \sum_k E_{d_k} + \sum_{lm} E_{vdW_{lm}}$. The functional forms of the contributing terms for a single interaction are as follows [25]:

$$E_b(d) = \frac{k_d}{2}(d - d_0)^2 \quad \text{for} \quad d < d_{cut}, \tag{1a}$$

$$E_a(\theta) = \frac{k_\theta}{2}(\theta - \theta_0)^2, \tag{1b}$$

$$E_d(\phi) = \frac{k_\phi}{2}[1 + \cos 2\phi], \tag{1c}$$

$$E_{vdW}(r) = D_0\left[\left(\frac{r_0}{r}\right)^{12} - 2\left(\frac{r_0}{r}\right)^6\right], \tag{1d}$$

where $k_d$ and $d_0$ are the spring constant of the bond length and the equilibrium bond distance, respectively; $k_\theta$ and $\theta_0$ are respectively the spring constant of the bond angle and the equilibrium bond angle; $k_\phi$ and $\phi$ are the spring constant of the dihedral angle and the dihedral angle, respectively. $D_0$ and $r_0$ are the Lennard-Jones parameters associated with the equilibrium well depth and the equilibrium distance, respectively. A potential cutoff of 1.25 nm is used in calculation of vdW interactions. The parameters of the force field are listed in Table 1.

Table 1. Parameters of the CG force field for C and P beads [25].

| Type of interaction | Parameters | C bead | P bead | C-P beads |
|---|---|---|---|---|
| Bond | $K_0(kcal/mol/Å^2)$ | 1610.24 | 194.61 | - |
| | $d_0(Å)$ | 6.03 | 4.02 | - |
| | $d_{cut}(Å)$ | 6.77 | 4.3 | - |
| Angle | $K_\theta(kcal/mol/Å^2)$ | 66148.01 | 794.89 | - |
| | $\theta_0(°)$ | 180 | 89.6 | - |
| Dihedral | $K_\phi(kcal/mol)$ | 14858.80 | 42.05 | - |
| vdW | $D_0(kcal/mol)$ | 10.68 | 1.34 | 2.8 |
| | $r_0(Å)$ | 9.45 | 6.53 | 7.71 |

An effective way to enhance the elastic properties of carbon nanotube reinforced polymer composites is the formation of covalent cross-links. Herein, we respectively choose 2CH$_2$ and EGDMA cross-links between polymer matrix and CNTs [27, 28], and polymer chains [29]. The atomistic and CG models of a 2CH$_2$ between a polymer chain and a CNT are respectively shown in Figs. 2a and 2b. The atomistic model of an EGDMA cross-link between two PMMA polymer chains and its corresponding CG counterpart are also illustrated in Figs. 2c and 2d. The cross-links are randomly added between nanotubes and polymer matrix, and between PMMA chains according to the following rules: (1) cross-links only connect beads that are placed within the equilibrium distance, (2) no cross-link connects a chain to itself, and (3) there are at least two beads between two sequential cross-links. The parameter of CG stretching potentials for $2CH_2$ and EGDMA cross-links taken from Ref. [30] are listed in Table 2. In this study, Accelrys Material Studio 7.0 is employed for conducting the simulations.

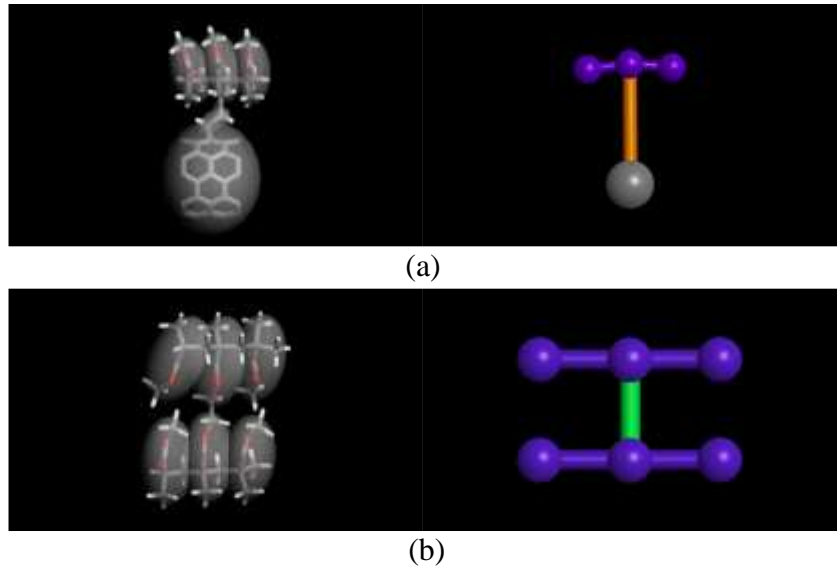

(a)

(b)

Fig. 2. (a) Atomistic and CG models of a $2CH_2$ cross-link between a PMMA polymer chain with three monomers and a (5, 5) CNT with five atomic rings, and (b) atomistic and CG models of an EGDMA cross-link between two PMMA polymer chains.

Table 2. Parameters of CG stretching potentials for $2CH_2$ and EGDMA cross-links [30].

| Type of cross-link | $K_0 (kcal/mol/Å^2)$ | $d_0 (Å)$ | $d_{cut} (Å)$ |
|---|---|---|---|
| $2CH_2$ | 142.71 | 9.487 | 10.11 |
| EGDMA | 150.206 | 6.21 | 7.47 |

## 3. Simulation method for calculating the J-integral

In this section, we extract the J-integral of reinforced amorphous polymer materials from CG simulations. Therefore, consider a double-edge-notched panel of thickness $B$ under tension shown in Fig. 3. Cracks of length $a$ on opposite sides of the panel are separated by a ligament of length 2b. The J-integral is then given by [31]

$$J = \frac{1}{2}\int_0^{P_{cr}} \left(\frac{\partial \Delta}{\partial a}\right)_P dP \qquad (2)$$

where $a$ is the initial length of notches as illustrated in Fig. 3.

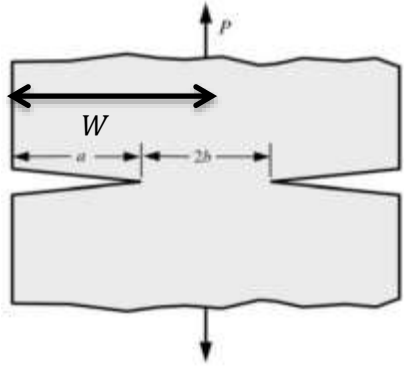

Fig. 3. Double-edge-notched tension panel.

The J-integral can be modified as [31]

$$J = \frac{K_I^2}{E'} + \frac{1}{2bB}\left[2\int_0^{\Delta_p} P d\Delta_p - P\Delta_p\right] \qquad (3)$$

where $E' = E$ for plane stress and $E' = E/(1-v^2)$ for plane strain. $K_I$ is the mode I stress intensity factor for elastic region given by [31]

$$K_I = \frac{P}{B\sqrt{W}}\sqrt{\frac{\pi a}{4W} \sec\left(\frac{\pi a}{2W}\right)}\left[1 - 0.025\left(\frac{a}{W}\right)^2 + 0.06\left(\frac{a}{W}\right)^4\right] \qquad (4)$$

in which $W = a + b$.

In order to measure $J$ from Eq. (3), it is necessary to determine the relationship between load and displacement. Herein, we demonstrate the utility of the above-mentioned approach in conjunction with the CG model for determining the magnitude of J-integral. First, a simulation unit cell with a size of 60×60×5 nm$^3$ and periodic boundary conditions that contains amorphous PMMA polymer and in-plane randomly distributed 10-nm long (5, 5) CNTs is constructed. Each polymer chain is composed of 100 repeated monomer units. The CNT weight fraction and the mass density of the CNT/PMMA composite are set to be 5 wt% and 1.1 g/cc, respectively. To obtain quasi-isotropic mechanical properties, a uniform probability distribution function is used to place the CNTs inside the polymer matrix. The CG model of the composite system contains 121116 beads that is equivalent to a full atomistic system with 2065020 atoms. To find a global minimum energy configuration of the system, a geometry optimization is initially performed using the conjugate-gradient method [32]. The isothermal–isobaric ensemble (NPT) ensemble at atmospheric pressure of 101 kPa is then used to gradually heat the system with a rate of 10 K/ps to room temperature of 298 K. The Andersen feedback thermostat [33] and the Berendsen barostat algorithm [34] are respectively used for the system temperature and pressure conversions. In the NPT simulations, the time step is set to be 0.1 fs and a geometry optimization is performed each 5 ps. The system is further allowed to equilibrate over an NPT ensemble at temperature of 298 K and pressure of 101 kPa for 5 ns with a time step of 10 fs. The NPT simulation is followed by an energy minimization. The process removes internal stresses in the composite system illustrated in Fig. 4a. Next, two center cracks, each of length 15 nm (i.e., $a/W = 0.5$) and width 0.5 nm, at opposite sides of the composite panel are created as shown in Fig. 4b. For the crack configuration, beads and bonds in the crack regions are deleted. An energy minimization is conducted again, and the system is relaxed over an NPT ensemble for 3 ns.

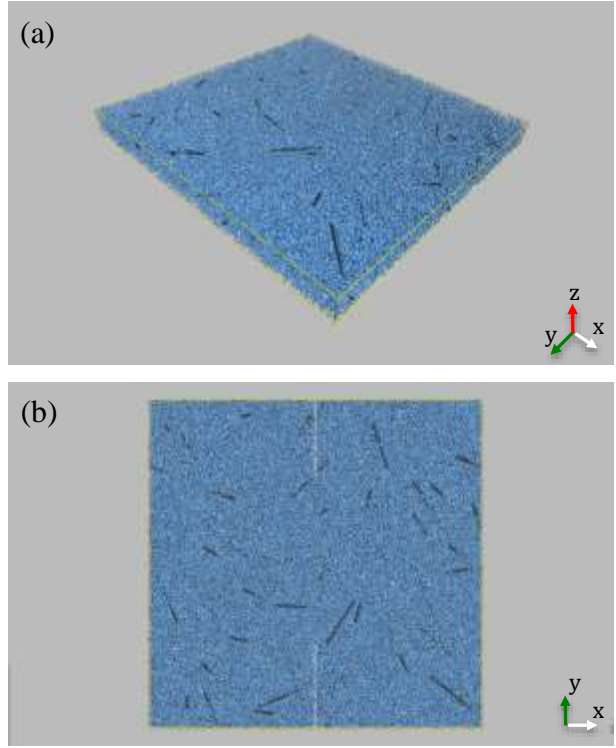

Fig. 4. An RVE of a CNT (5 wt%)/PMMA composite with a size of 60×60×5 nm$^3$ and 15-nm long initial notches at two sides in the *y*-direction. The length and diameter of CNTs are 10 and 0.68 nm, respectively: (a) perspective view, and (b) top view. The CG model of the composite system contains 121116 beads that is equivalent to a full atomistic system with 2065020 atoms.

After the preparation of the system, the constant-strain minimization method is applied to the equilibrated system to measure the material properties of the composite [3, 35]. A small tensile strain of 0.02% is applied to the periodic structure shown in Fig. 4b in the perpendicular direction to the direction of cracks (i.e., the *x*-direction). The application of the static strain is accomplished by uniformly expanding the dimensions of the simulation cell in the loading direction and re-scaling the new coordinates of the atoms to fit within the new dimensions. After each increment of the applied strain, the potential energy of the structure is re-minimized keeping the lattice parameters (i.e., simulation box sizes and angles) fixed. During the static deformation, the pressure in *y*- and *z*-directions is kept at atmospheric pressure by controlling the lateral dimensions. This process is repeated for a series of strains from which the variation of potential energy of the system versus displacement can be obtained as presented in Fig. 5.

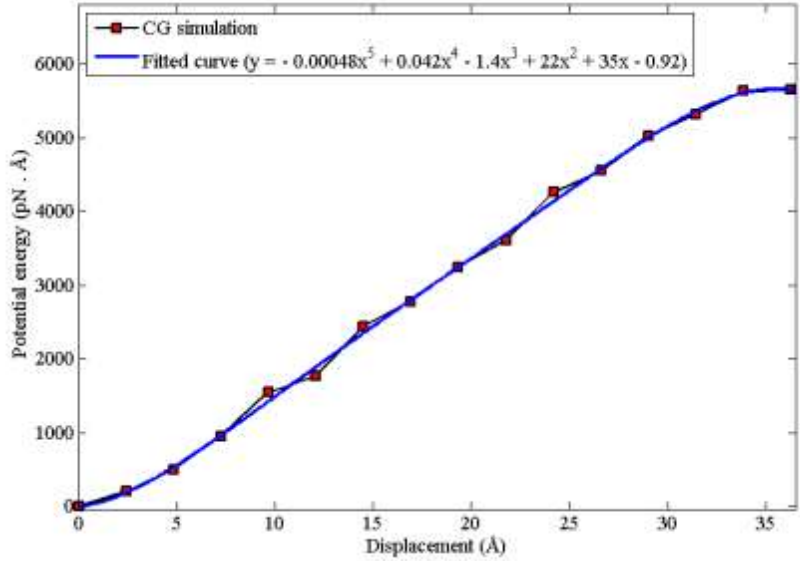

Fig. 5. Variation of potential energy versus displacement of the CNT/PMMA composite. The length and weight fraction of nanotubes are 10 nm and 5 wt%, respectively.

The load-displacement curve is then obtained from the first derivative of potential energy with respect to displacement as $P = \frac{\partial U}{\partial \Delta}\big|_\Delta$. Fig. 6 presents the load-displacement curve of the double-edge-notched composite panel in Fig. 4b. The initiation and propagation of the cracks in the reinforced polymer system at different load levels (see Figs. 7 a-d) reveal that, as physically expected, the crack growth occurs along the ligament between two initial notches. Hence, mode I fracture is dominant and the J-integral can be extracted from the load-displacement curve presented in Fig. 6. Using Eq. (3), then J-integral of the CNT (5 wt%)/PMMA composite specimen with a size of 60×60×5 nm³ and a 15-nm long initial notches ($a/W = 0.5$) is calculated to be $J = 0.0068\ J/m^2$.

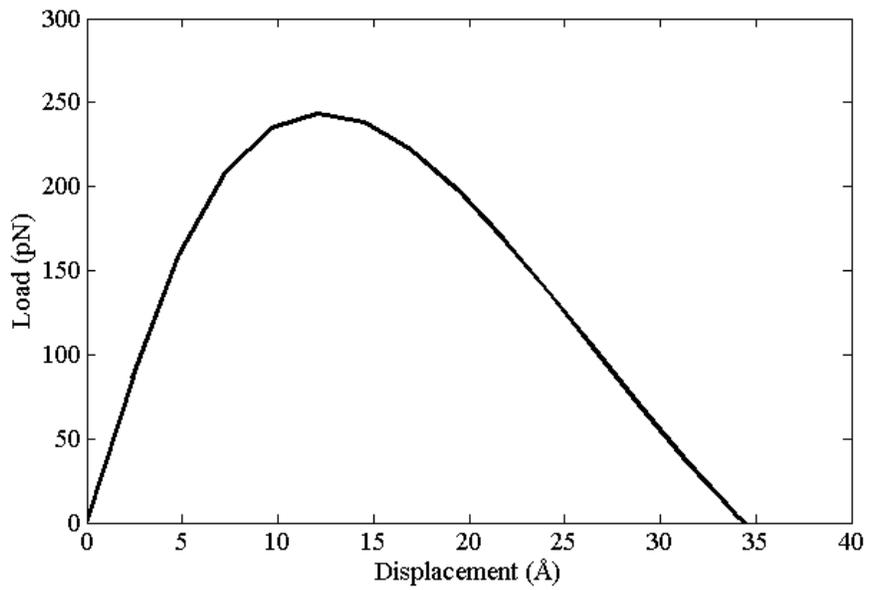
Fig. 6. Load-displacement curve of the CNT(5 wt%)/PMMA composite.

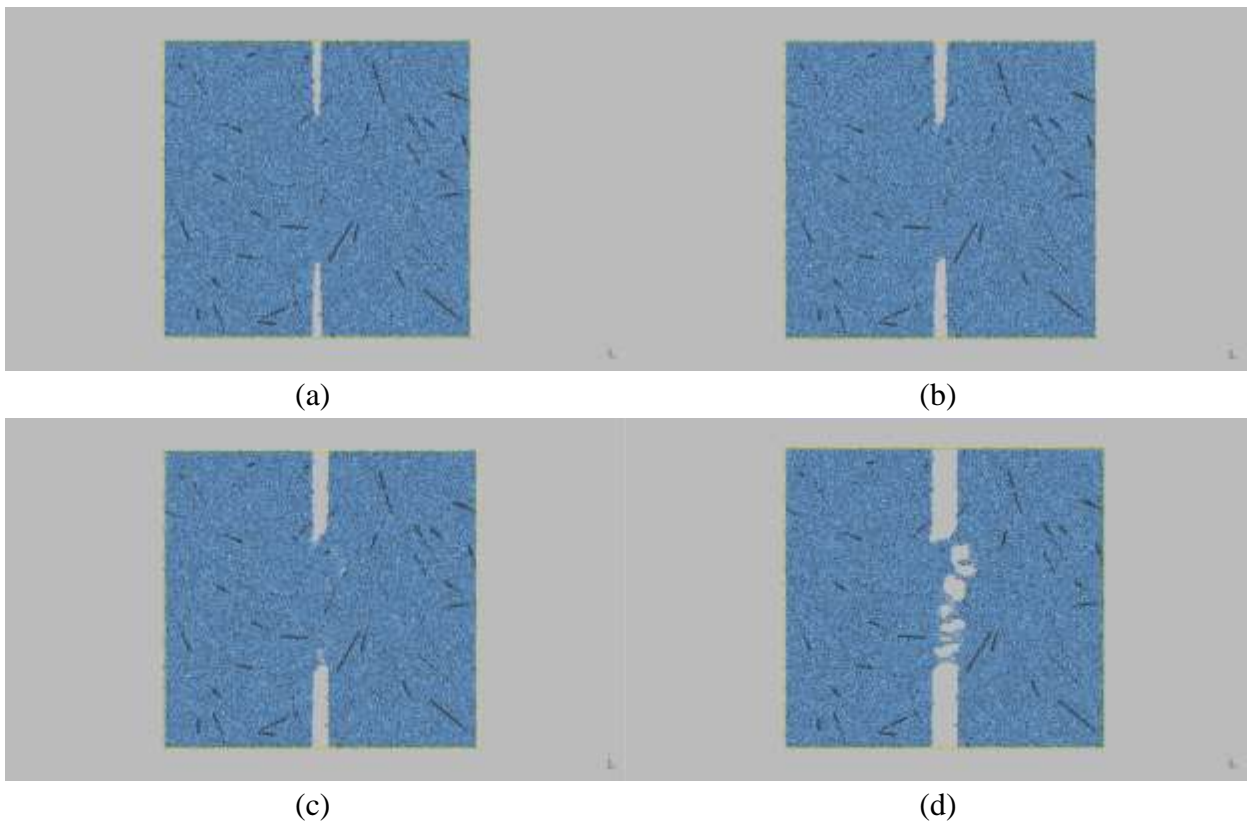
Fig. 7. Fracture propagation in the CNT (5 wt%)/PMMA composites: (a) snapshot at $\Delta = 10$ Å, (b) snapshot at $\Delta = 12$ Å, (c) snapshot at $\Delta = 15$ Å, and (d) snapshot at $\Delta = 30$ Å.

## 4. Results and discussion

### 4.1 Influence of CNT weight fraction

We first study the influence of the CNT weight fraction on the load-displacement response. In the following simulations, the RVE size, the size of nanotubes, the mass density of polymer composite and the length of initial notches are the same as described in Fig. 4. CNT fibers are in-plane randomly distributed and their weight fraction varies from 0 to 10 wt%. Fig. 8 presents the load-displacement behavior of PMMA polymer and its CNT reinforced composites under the static tensile loading predicted by the CG model. The simulation results reveal that the polymer matrix exhibits a more ductile behavior in the presence of nanotube reinforcements. Following the method described in section 3, the J-integral of the composite is calculated from the load-displacement curves as listed in Table 3. From Table 3, the magnitude of J-integral of a pure PMMA polymer RVE with a size of 60×60×5 nm$^3$ and 15-nm long initial notches (i.e., $a/W = 0.5$) is calculated to be 0.0032 $J/m^2$. From Table 3, the J-integral of CNT/PMMA composite increases from 0.0032 $J/m^2$ to 0.0068 $J/m^2$ with an increase in the weight fraction of nanotubes from 0 to 5 wt%, showing a percentage increase of around 112.5 %. The J-integral further increases to 0.0093 $J/m^2$ by increasing the weight fraction of CNTs to 10 wt%. The simulation results indicate that short CNT fibers with the length-to-diameter ($L/D$) aspect ratio of 14.7 and weight fractions of 10 wt% significantly raise the J-integral of a PMMA polymer matrix by as much as a factor of three. This trend can be interpreted in the sense that nanotube reinforcements enhance the resistance of the polymer matrix to opening a crack, which is consistent with experimental observations reported in the literature [36].

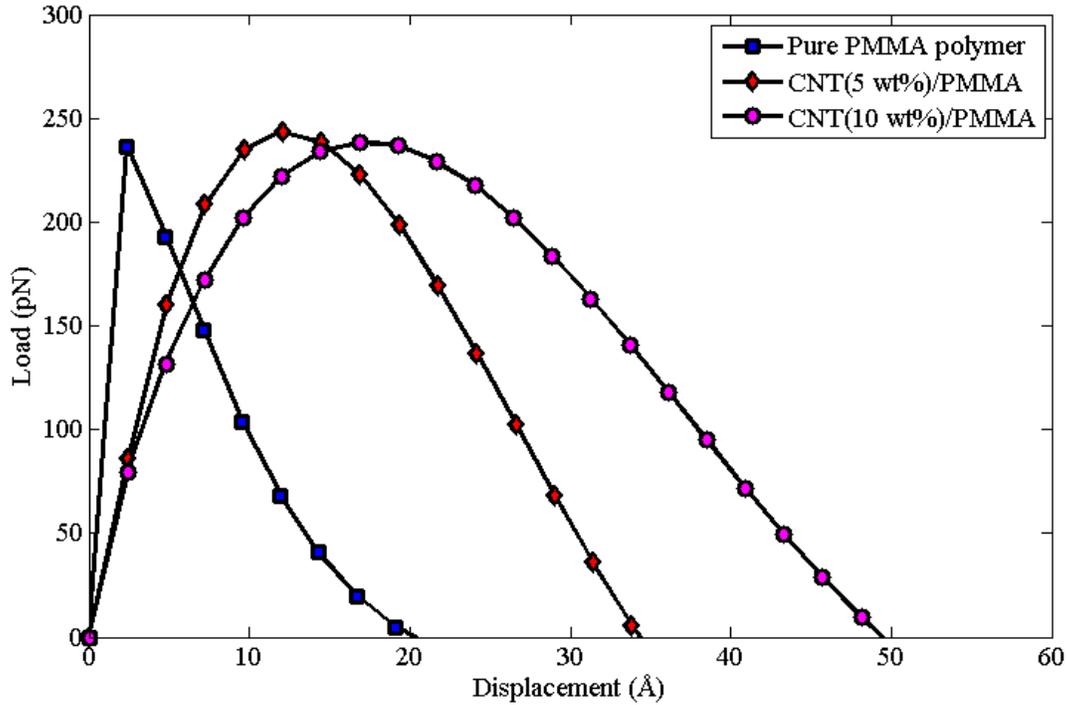

Fig. 8. Effect of the CNT weight fraction on the load-displacement curve of CNT/PMMA composites.

Table 3. J-integral of CNT/PMMA composites without cross-links. The aspect ratio of $a/W$ is set to be 0.5 and the RVE size is 60×60×5 nm$^3$.

| CNT weight fraction (%) | J-integral ($J/m^2$) |
|---|---|
| **0** | 0.0032 |
| **5** | 0.0068 |
| **10** | 0.0093 |

### 4.2 Influence of cross-links

In order to further study the fracture behavior of CNT/PMMA composites, the effect of cross-links between the polymer matrix and nanotubes, and polymer chains on the amount of J-integral is investigated in the following simulations. The RVE size and the weight fraction of 10-nm long (5, 5) CNTs are 60×60×5 nm$^3$ and 10 wt%, respectively. The length of initial notches is the same as previous simulations (i.e, $a = 15\ nm$). After the preparation of the composite systems as described in section 3, 2CH$_2$ and EGDMA cross-links are respectively formed between PMMA polymer and nanotubes, and polymer chains as illustrated in Fig. 9. In order to equilibrate the system, a NPT simulation is performed at room temperature of 298 K and atmospheric pressure of 101 kPa for 50 ns ps with a time step of 1 fs. The total energy of the system is re-minimized and another NPT simulation is carried out at the same pressure and temperature conditions for 3 ns with a time step of 10 fs. The NPT simulation is followed by a further energy minimization to remove internal stresses in the system. The constant-strain minimization method is then applied to the system to obtain the load-displacement curve from which the J-integral fracture toughness is measured.

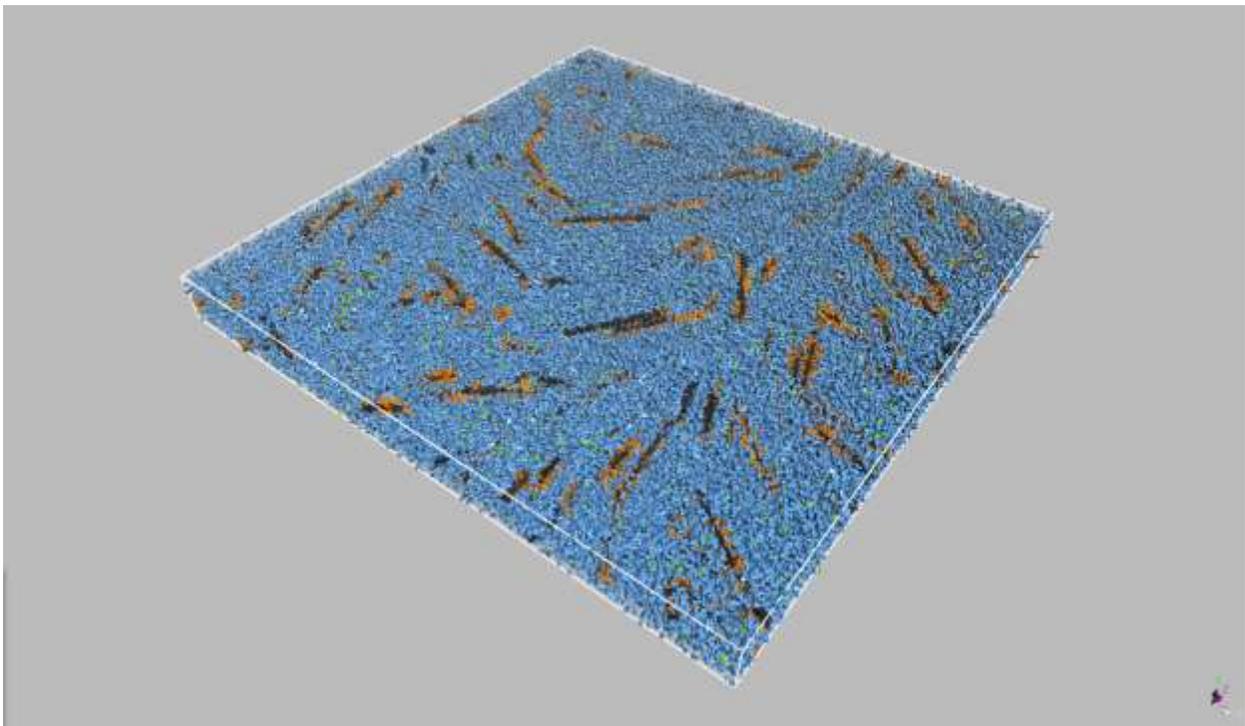

Fig. 9. A CNT (10 wt%)/PMMA composite with EGDMA cross-links between polymer chains and 2CH$_2$ cross-links between polymer chains and CNTs. The RVE size is 60×60×5 nm$^3$ and the 10-nm long (5, 5) CNT reinforcements are randomly distributed in plane. Cross-links between C-P beads and P-P beads are specified with orange and green connectors, respectively. The RVE contains 121120 beads which are equivalent to 2065091 atoms.

First, the effect of 2CH$_2$ cross-links between nanotubes and polymer matrix on the fracture toughness of the CNT(10 wt%)/PMMA composites is investigated. The J-integral of the CNT (10 wt%)/PMMA composite versus the mole fraction of 2CH$_2$ cross-links is presented in Fig. 10. The cross-link mole fraction is defined as the ratio of the amount of mole of cross-links to the total amount of moles of the composite system. From Fig. 10, the J-integral increases from 0.0093 to 0.0110 J/m² with an increase in the mole fraction of the cross-links from 0 to 2%, indicating a percentage increase of 18%. By further increasing the cross-link mole fraction to 6 and 8%, the J-integral of the composite is calculated to be 0.0142 and 0.0168 $J/m^2$, respectively. The simulation results indicate that 2CH$_2$ cross-links with mole fractions of 6 and 8% respectively enhance the fracture resistance of the CNT/PMMA composite as much as 53 and 81% compared to the composite without cross-links. The simulations also reveal that the J-integral of the polymer composites with functionalized CNTs with a weight fraction of 10 wt% and cross-link mole fraction of 8% is more than 5 time greater than that of the pure PMMA polymer. The significant increase in the J-integral fracture toughness occurs because the covalent cross-links between nanotube fibers and polymer chains strengthen the interfacial region between nanofillers and polymer matrix, which in turn improves the resistance of the composites to crack growth.

The influence of EGDMA cross-links between polymer chains on the J-integral of the composites is presented in Fig. 11. In the simulations, the mole fraction of 2CH$_2$ cross-links between nanotubes and polymer chains is set to be 8%, while the mole fraction of EGDMA cross-links varies from 0 to 8%. With an increase in the mole fraction of EGDMA cross-links from 0 to 2%, the J-integral of the CNT/PMMA composite is raised from 0.0168 to 0.0395 $J/m^2$, showing a percentage increase of 135%. The J-integral of the composite with both cross links of 2CH$_2$ and EGDMA reaches 0.096 $J/m^2$ at the mole fractions of 8%. The simulation results demonstrate that the J-integral of the CNT(10 wt%)/PMMA with 8% mole fraction of 2CH$_2$ and EGDMA cross-links is respectively 10 and 30 times greater than those of CNT/PMMA composite without the covalent cross-links and pure polymer. It implies that the formation of covalent cross links in the polymer composite remarkably improves the fracture toughness of randomly distributed CNT/PMMA composites.

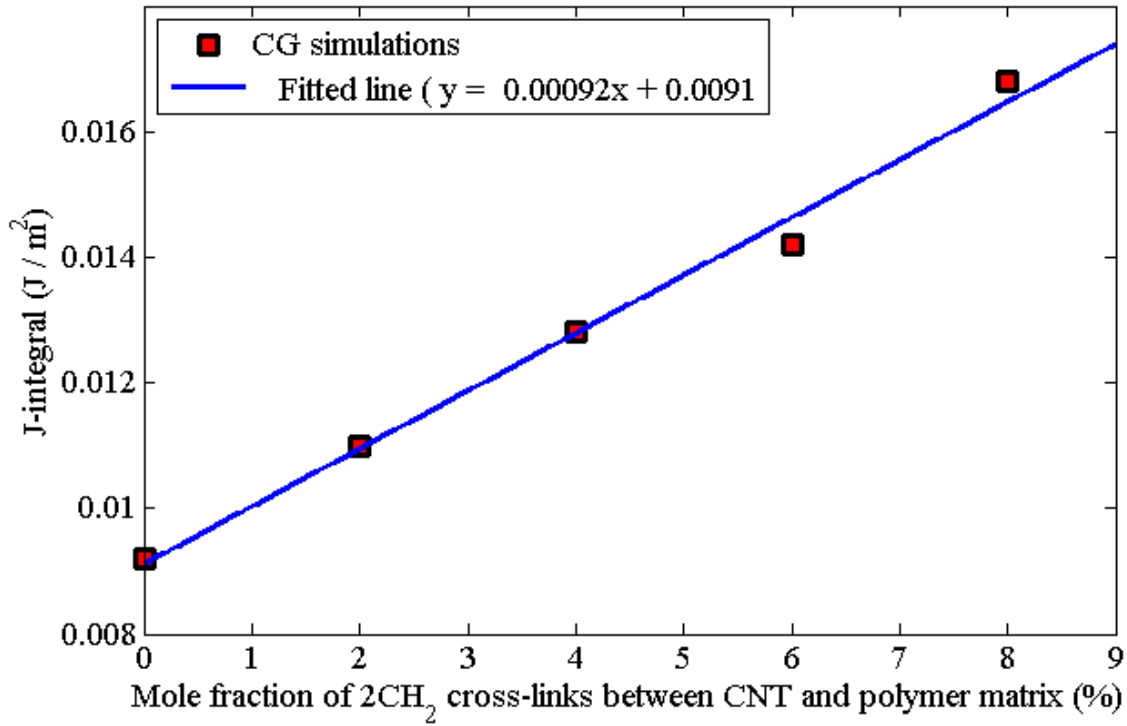

Fig. 10. Effect of mole fraction of 2CH$_2$ cross-links between CNTs and polymer matrix on the magnitude of J-integral of CNT/PMMA composites. The RVE size is 60×60×5 nm$^3$, the CNT weight fraction is 10 wt% and the aspect ratio of $a/W$ is set to be 0.5.

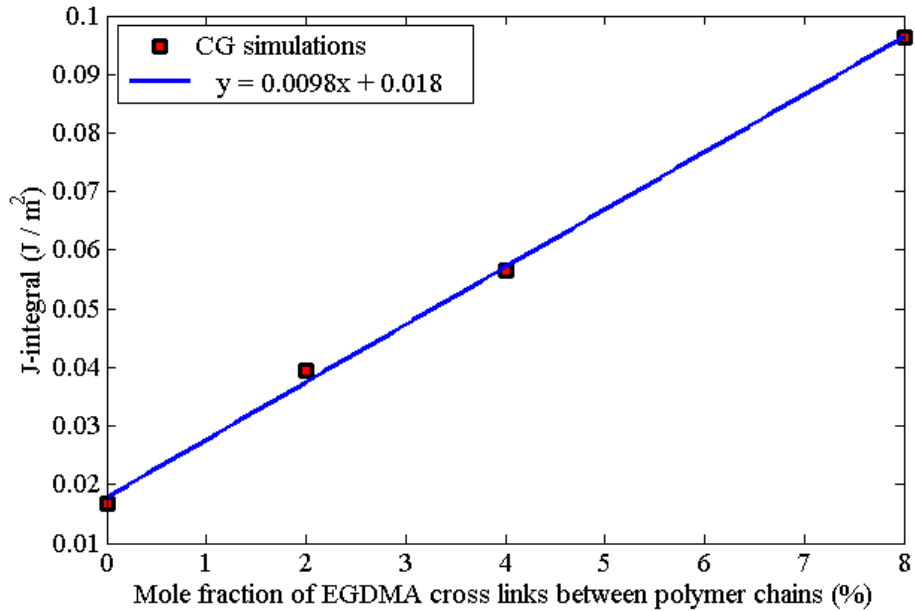

Fig. 11. Effect of mole fraction of both EGDMA cross-links between polymer chains on the magnitude of J-integral of CNT/PMMA composites. The mole fraction of 2CH$_2$ cross-links between polymer chains and CNTs is set to be 8%. The RVE size is 60×60×5 nm$^3$, the CNT weight fraction is 10 wt% and the aspect ratio of $a/W$ is set to be 0.5.

### 4.3 Influence of notch length

After determining the influence of covalent cross-links on the fracture behavior of CNT/PMMA composites, the effect of the length of initial notches on the fracture toughness of the nanocomposites is investigated. The following simulations enable to further evaluate the reliability of the proposed method in predicting the J-integral of the polymer composites. Fig. 12 presents the effect of $a/W$ ratio on the load-displacement response of CNT(10 wt%)/PMMA composites with 2CH$_2$ cross-links between nanotube fibers and polymer matrix under the static tensile test. In the simulations, the RVE size is 60×60×5 nm$^3$, the mole fraction of 2CH$_2$ cross-links is 4%, and the $a/W$ ratio varies from 0.3 to 0.6. To adjust the value of the $a/W$ ratio, the length of initial notches illustrated in Fig. 4b differs from $a = 9$ to 18 nm, while the width of the RVE is fixed at 60 nm (i.e., $2W = 60\ nm$). Based on the method proposed in section 3, the J-integral of the composites can be calculated from the load-displacement curves as listed in Table

4. The value of J-integral is obtained to be 0.0392 $J/m^2$ at $a/w = 0.2$. The magnitude of J-integral reported in the literature for a gold crystal [19] with a sample size of ~ 65×65×2.3 nm$^3$ (similar to the RVE size of the polymer composite) and the same $a/w$ ratio of 0.2 is 0.076 $J/m^2$. It demonstrate that the value of J-integral predicted by the present method for the reinforced polymer composite is rationally smaller than that of measured for gold. The magnitude of J-integral of the CNT/PMMA composite decreases from 0.0288 to 0.0183 $J/m^2$ with an increase in the $a/w$ ratio from 0.3 to 0.4. The J-integral drops to 0.0128 and 0.0111 $J/m^2$ by increasing the $a/w$ ratio to 0.5 and 0.6. The simulation results reveal that the fracture toughness of the composite sample significantly reduces as much as 72% with increasing the $a/w$ ratio from 0.2 to 0.6. Physically, it implies that the energy required for the propagation of the initial cracks through the entire specimen width decreases with increasing the length of initial notches, which is consistent with experimental and numerical data [19, 37].

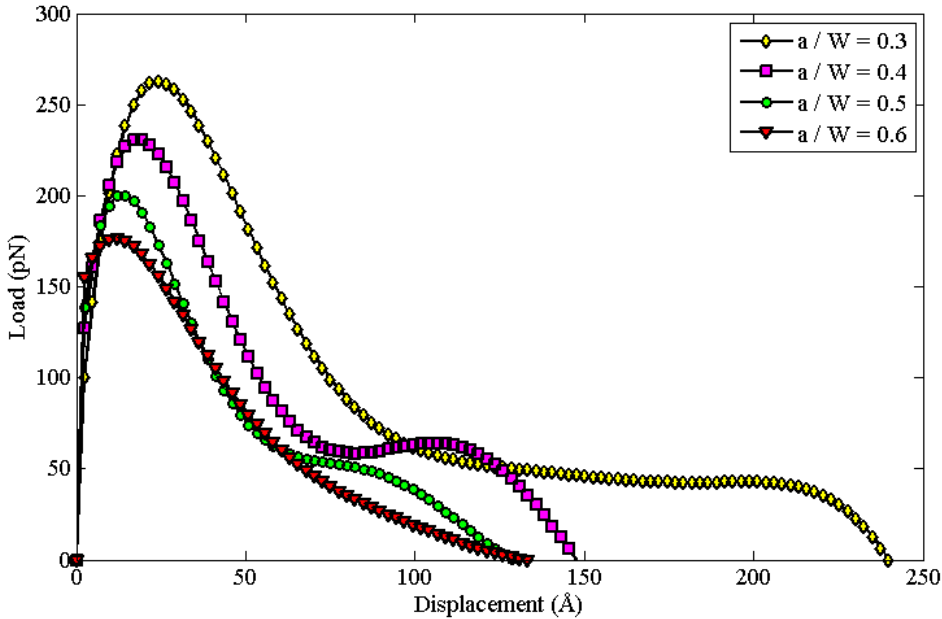

Fig. 12. Effect of $a/W$ ratio on the load-displacement response of CNT(10 wt%)/PMMA composites with 2CH$_2$ cross-links between nanotubes and polymer matrix. The RVE size is 60×60×5 nm$^3$, the CNT weight fraction is 10 wt%, and the mole fraction of 2CH$_2$ cross-links is 4%.

Table 4. Effect of $a/W$ ratio on the J-integral fracture toughness of CNT(10 wt%)/PMMA with 2CH$_2$ cross-links between nanotubes and polymer matrix. The RVE size is 60×60×5 nm$^3$, the CNT weight fraction is 10 wt%, and the mole fraction of 2CH$_2$ cross-links is 4%.

| $a/W$ | J-integral ($J/m^2$) |
|---|---|
| 0.2 | 0.0392 |
| 0.3 | 0.0288 |
| 0.4 | 0.0183 |
| 0.5 | 0.0128 |
| 0.6 | 0.0111 |

**4.4 Influence of RVE size**

Finally, we explore the effect of the RVE size on the magnitude of J-integral of CNT/PMMA composites with 2CH$_2$ cross-links between nanotubes and polymer matrix in Table 5. In the simulations, the $a/w$ ratio is fixed to be 0.5, and the RVE side lengths vary from 60 to 100 nm. The RVEs contain between 121116 and 336433 beads, which are equivalent to 2065020 to 5046500 atoms. The CNT weight fraction and the mole fraction of 2CH$_2$ are set to be 10 wt% and 4%, respectively. From Table 5, the J-integral of the cross-linked CNT(10 wt%)/PMMA composite increases from 0.0128 to 0.0225 $J/m^2$ with an increase in the RVE side length from 60 to 80 nm. The simulation results indicate a percentage increase of about 76% in the J-integral fracture toughness of the composite by increasing the length of ligament ($= 2b$), which is shown in Fig. 3, from 30 to 40 nm. The J-integral is further raised to 0.0372 $J/m^2$ with increasing the RVE side length to 100 nm, showing a percentage increase of 190% compared to the specimen with a side length of 60 nm. The simulations show that the energy release rate increases by increasing the length of ligament between to initial notches. In other words, the fracture energy required to open the crack (the material resistance against fracture) enhances when the RVE size and the length of initial notches are proportionally enlarged. This observation is in agreement with experimental data [38, 39].

Table 5. Effect of the RVE size on the magnitude of J-integral. The CNT weight fraction is 10 wt%, the mole fraction of 2CH$_2$ cross-links is equal to 4%, and the aspect ratio of $a/W$ is set to be 0.5.

| RVE size (nm$^3$) | Number of beads (equivalent number of atoms) | Length of ligament (2b) (nm) | J-integral ($J/m^2$) |
|---|---|---|---|
| **60×60×5** | 121116 (2065020) | 30 | 0.0128 |
| **80×80×5** | 215318 (3229760) | 40 | 0.0225 |
| **100×100×5** | 336433 (5046500) | 50 | 0.0372 |

## 5. Conclusions

A J-integral calculation method based on CG simulations is presented to study the fracture behavior of CNT reinforced polymer composites under tension. In the simulation method, the J-integral is obtained from the load displacement curve of a single double-edge-notched specimen. The proposed approach provides an effective methodology for predicting the J-integral of polymers and their composites without requiring knowledge about the crack tip position or the continuum fields around the tip. The CG model beyond the capacity of conventional molecular simulations enables to substantially increase the accessible length-scales compared to full atomistic systems. The effects of the nanotube weight fraction and cross-links between the polymer matrix and CNTs, and polymer chains on the fracture behavior and the J-integral fracture toughness of the CNT/PMMA composites are investigated. The dependence of the J-integral on the crack length and the RVE size is discussed and it is shown that the simulation observations are in consistent with experimental and numerical data available in the literature. The simulation results show that 10-nm long (5, 5) CNT reinforcements with a weight fraction of 10 wt% significantly enhance the J-integral of PMMA polymer matrix as much as 3 times. The simulations also reveal that functionalized CNTs with a weight fraction of 10 wt% induce a considerable increase in the magnitude of J-integral of the CNT/PMMA composites as much as 5 times in the presence of 2CH$_2$ cross-links between nanotubes and polymer matrix with a mole fraction 8%. Based on the simulations, the formation of 2$CH_2$ between nanotubes and polymer matrix and EGDMA cross-links between polymer chains with mole fractions of 8% increases the J-integral of the CNT(10 wt%)/PMMA composite as much as 10 and 30 times compared to the composite without covalent cross-links and pure PMMA polymer, respectively.

Although the proposed method promises an efficient method for predicting the J-integral of complex structures at the nanoscale, further investigations are evident in the future. Our method needs to be extended to study the effect strain rate on the magnitude of J-integral of polymer composites. An expansion of the method is also required to study the fracture behavior of polymer systems subjected to finite temperature conditions.

**Acknowledgments**

The authors thank the support of the European Research Council-Consolidator Grant (ERC-CoG) under grant "Computational Modeling and Design of Lithium-ion Batteries (COMBAT)".

**Figure captions**

Fig. 1. (a) Two monomers of a PMMA polymer chain and its CG model made of two P beads, and (b) a (5, 5) CNT with 10 rings of carbon atoms and its CG model made of two C beads.

Fig. 2. (a) Atomistic and CG models of a 2CH$_2$ cross-link between a PMMA polymer chain with three monomers and a (5, 5) CNT with five atomic rings, and (b) atomistic and CG models of an EGDMA cross-link between two PMMA polymer chains.

Fig. 3. Double-edge-notched tension panel.

Fig. 4. An RVE of a CNT (5 wt%)/PMMA composite with a size of 60×60×5 nm$^3$ and 15-nm long initial notches at two sides in the *y*-direction. The length and diameter of CNTs are 10 and 0.68 nm, respectively: (a) perspective view, and (b) top view. The CG model of the composite system contains 121116 beads that is equivalent to a full atomistic system with 2065020 atoms.

Fig. 5. Variation of potential energy versus displacement of the CNT/PMMA composite. The length and weight fraction of nanotubes are 10 nm and 5 wt%, respectively.

Fig. 6. Load-displacement curve of the CNT(5 wt%)/PMMA composite.

Fig. 7. Fracture propagation in the CNT (5 wt%)/PMMA composites: (a) snapshot at $\Delta = 10$ Å, (b) snapshot at $\Delta = 12$ Å, (c) snapshot at $\Delta = 15$ Å, and (d) snapshot at $\Delta = 30$ Å.

Fig. 8. Effect of the CNT weight fraction on the load-displacement curve of CNT/PMMA composites.

Fig. 9. A CNT (10 wt%)/PMMA composite with EGDMA cross-links between polymer chains and 2CH$_2$ cross-links between polymer chains and CNTs. The RVE size is 60×60×5 nm$^3$ and the 10-nm long (5, 5) CNT reinforcements are randomly distributed in plane. Cross-links between C-P beads and P-P beads are specified with orange and green connectors, respectively. The RVE contains 121120 beads which are equivalent to 2065091 atoms.

Fig. 10. Effect of mole fraction of 2CH$_2$ cross-links between CNTs and polymer matrix on the magnitude of J-integral of CNT/PMMA composites. The RVE size is 60×60×5 nm$^3$, the CNT weight fraction is 10 wt% and the aspect ratio of $a/W$ is set to be 0.5.

Fig. 11. Effect of mole fraction of both EGDMA cross-links between polymer chains on the magnitude of J-integral of CNT/PMMA composites. The mole fraction of 2CH$_2$ cross-links between polymer chains and CNTs is set to be 8%. The RVE size is 60×60×5 nm$^3$, the CNT weight fraction is 10 wt% and the aspect ratio of $a/W$ is set to be 0.5.

Fig. 12. Effect of $a/W$ ratio on the load-displacement response of CNT(10 wt%)/PMMA composites with 2CH$_2$ cross-links between nanotubes and polymer matrix. The RVE size is 60×60×5 nm$^3$, the CNT weight fraction is 10 wt%, and the mole fraction of 2CH$_2$ cross-links is 4%.

**Table captions**

Table 1. Parameters of the CG force field for C and P beads [24].

Table 2. Parameters of CG stretching potentials for 2CH$_2$ and EGDMA cross-links [29].

Table 3. J-integral of CNT/PMMA composites without cross-links. The aspect ratio of $a/W$ is set to be 0.5 and the RVE size is 60×60×5 nm$^3$.

Table 4. Effect of $a/W$ ratio on the J-integral fracture toughness of CNT(10 wt%)/PMMA with 2CH$_2$ cross-links between nanotubes and polymer matrix. The RVE size is 60×60×5 nm$^3$, the CNT weight fraction is 10 wt%, and the mole fraction of 2CH$_2$ cross-links is 4%.

Table 5. Effect of the RVE size on the magnitude of J-integral. The CNT weight fraction is 10 wt%, the mole fraction of 2CH$_2$ cross-links is equal to 4%, and the aspect ratio of $a/W$ is set to be 0.5.